# Twin Estimates of the Effects of Prenatal Environment, Child Biology, and Parental Bias on Sex Differences in Early Age Mortality[1]

Roland Pongou[2]

This version: May 2010

**Abstract:** Sex differences in early age mortality have been explained in prior literature by differences in biological make-up and gender discrimination in the allocation of household resources. Studies estimating the effects of these factors have generally assumed that offspring sex ratio is random, which is implausible in view of recent evidence that the sex of a child is partly determined by prenatal environmental factors. These factors may also affect child health and survival in *utero* or after birth, which implies that conventional approaches to explaining sex differences in mortality are likely to yield biased estimates. We propose a methodology for decomposing these differences into the effects of prenatal environment, child biology, and parental preferences. Using a large sample of twins, we compare mortality rates in male-female twin pairs in India, a region known for discriminating against daughters, and sub-Saharan Africa, a region where sons and daughters are thought to be valued by their parents about equally. We find that: (1) prenatal environment positively affects the mortality of male children; (2) biological make-up of the latter contributes to their excess mortality, but its effect has been previously overestimated; and (3) parental discrimination against female children in India negatively affects their survival; but failure to control for the effects of prenatal and biological factors leads conventional approaches to underestimating its effect by 237 percent during infancy, and 44 percent during childhood.

---

[1]I thank Anna Aizer, Ken Chay, Andrew Foster, Emily Oster, Mark Pitt, Yona Rubinstein, Olumide Taiwo for useful comments on earlier versions of this paper.

[2]Dept. of Economics, Brown University, Providence, RI 02912, U.S.A.; Roland_Pongou@brown.edu

# 1 Introduction

We investigate the origins of sex imbalance in early age mortality. It has long been observed that female children have a better survival chance than male children (Graunt (1662)). However, in some Asian countries, the former lose their advantage before the age of five (see Figure 1).[3] These facts have been respectively explained by female biological advantage[4], and pro-male bias in investment by parents and other care-takers.[5]

A common assumption made in studies testing the biological and economic theories of sex differences in mortality is that offspring sex ratio is randomly assigned across and within families.[6] Recent literature however reveals that this is not the case (e.g., James (1998), Almond and Mazumber (2009), Almond and Edlund (2007)).[7] According to James (1998), offspring sex ratio is partly determined by parental circumstances and levels of hormones at the time of conception. The likelihood of bearing a son is increased by high concentrations of testosterone and estrogen, and the likelihood of bearing a daughter is increased by high concentrations of gonadotrophins and progesterone. Levels of parental hormones are in turn related to parental stresses, illnesses and occupations (James (1998)). In more recent studies, Almond and Edlund (2007) and Almond and Mazumber (2009) find social class and maternal fasting during Ramadan, respectively, to affect sex ratio at birth. To the extent that these prenatal factors affect a baby's health and survival *in utero* or after birth, we believe that conventional methodological approaches adopted in previous studies produce biased estimates of the effects of biology and gender discrimination on sex differences in mortality. In this paper, we seek to overcome this bias. More precisely, we propose a methodology for decomposing the sex gap in mortality into the distinct effects of prenatal environment[8], child biology, and female discrimination.

---

[3] Figure 1 shows that mortality rates are greater for males than females during the first year of life in sub-Saharan Africa and India; however, between the first and fifth birthdays, while female children continue to die less frequently than their male counterparts in sub-Saharan Africa, the pattern is reversed in India.

[4] It is argued that male children are biologically weaker, and therefore are more susceptible to premature death than their female counterparts (e.g., Waldron (1983)).

[5] There is a large body of literature on the neglect of female children in South and East Asia: Sen (1984, 1989, 1990b, 1992), Lin and Luoh (2008), Lin, Luoh and Qian (2009), Klasen and Wink (2002), Coale and Banister (1994), Behrman (1988), Kynch and Sen (1983), Alderman and Gertler (1997), Basu (1992), Basu (1989), Chen, Huq and D'Souza (1981), Borooah (2004), Hazarika (2000), Pande (2003), Oster (2009), Sen and Sengupta (1983), Croll (2001), Preston and Weed (1976)). From an economic point of view, it is argued that higher parental investment in sons responds to differential labor market returns by sex (e.g., Rosenzweig and Schultz (1982), Sen (1990a)). This view is consistent with studies showing that female labor market participation and higher female income and education decrease the relative mortality of girls (Rose (1999), Qian (2008), World Bank (2001), Drèze and Sen (1996)).

[6] Offspring sex ratio, which is the ratio of male children to female children born to a parent, should not be confused with population sex ratio (at birth), which is the ratio of male children to female children born in a population. The world population sex ratio is estimated by demographers to be about 1.05, but offspring sex ratio varies widely across parents.

[7] This literature is briefly reviewed in Section 2.

## 1.1 An overview of the methodology and results

Our identification strategy relies on comparing the sex difference in mortality rates for "all twins" with "male-female twin pairs".[9] We posit that the sex difference in mortality rates for all twins (denoted A) is the additive effects of child biology, prenatal environment, and parental discrimination.[10][11] However, the sex difference in mortality rates for male-female twin pairs (denoted B) is the additive effects of biology and parental discrimination only.[12] Subtracting B from A thus yields the effects of prenatal environment.

To estimate the distinct effects of biology and parental discrimination, we first assume that there are two types of societies: one that discriminates against female children (called D), and one that is non-discriminatory (called ND). In ND, sons and daughters are valued by their parents and other care-takers about equally. Therefore, while in D, B represents the additive effects of biology and parental discrimination, in ND, B only represents the effects of biology. At this point, our methodology shows how the distinct effects of biology and prenatal environment are estimated in ND, the effect of discrimination in that society being zero by definition.

---

[8] By prenatal environment, we mean factors that are external to a child and that occur before conception. These factors might be pure environmental hazards such as parental exposure to chemicals, or medical factors such as parental illnesses. We are therefore not concerned with intrauterine environment in this study, but we discuss its potential effect in Section 6.

[9] All twins include same-sex twins (male-male twins and female-female twins) and mixed-sex twins (male-female twins).

[10] Additivity is consistent with models generally used by biologists and geneticists to disentangle the effects of genetic and environmental factors on health outcomes (e.g., Evans et al. 2002, Neale and Cardon 1992); see also Fowler et al. (2008) for a study on the role of genetic factors in political participation. These studies have found little supportive evidence for interactions between environment and biology.

[11] For ease in the interpretation of results, A is estimated using a linear probability model (LPM). Using a Probit model changes little to the results.

[12] B is estimated (from the same sample of twins as A) using a twin fixed effect (FE) LPM. Note that this latter model automatically controls for the effect of prenatal environment (here understood as *preconception* environment). Also, it is easy to notice that estimating a twin FE LPM in the sample of "all twins" is equivalent to estimating an LPM in the sub-sample of "male-female twin pairs."

Next, we separate out the distinct effects of biology and parental discrimination in D. We assume that biology has the same effects in D as in ND. And given that we know the effect of biology in ND (which is B in ND), and the additive effects of biology and parental discrimination in D (which is B in D), we derive the effects of discrimination in D by subtracting B in ND from B in D. Therefore, the effects of parental bias are obtained by comparing sex differences in mortality rates (estimated from the sample of male-female twin pairs) in the biased and unbiased populations. This completes our decomposition exercise.

We apply this decomposition methodology to samples of twins extracted from Demographic and Health Surveys. These are nationally representative cross-sectional surveys conducted in most developing countries. For our purposes, we use data from sub-Saharan Africa and India. Numerous studies have shown that sons and daughters are valued about equally in the former region, while daughters are discriminated against in the latter, as in several other countries of South and East Asia.[13] When it comes to child mortality, comparing sub-Saharan Africa with India is also pertinent because children in those regions die from similar diseases (Black et al. (2003)). This also motivates our assumption that biology has the same effects in both regions.[14]

In decomposing sex differences in mortality, we distinguish two main periods of child development: infancy, which is the period from birth to 12 months, and childhood (12-60 months). Infancy is further divided up into two periods: the neonatal period (birth-1 month), and the postneonatal period (1-12 months). We find that the male-female difference in infant mortality rates estimated from the sample of all twins (that is, the additive effects of biology, prenatal environment, and parental discrimination (A)) is 45 per thousand points in sub-Saharan Africa and 27 per thousand points in India. The male-female difference in infant mortality rates estimated from the sample of male-female twin pairs (that is, the additive effects of biology and parental discrimination (B)) is 27 and -10 per thousand points, respectively, in these regions.[15] It follows from our decomposition methodology that in sub-Saharan Africa, prenatal environment, biology and parental discrimination raise the mortality of male twins by 18, 27 and 0 per thousand points, respectively. In India, prenatal environment and biology raise the mortality of male twins by 37 and 27 per thousand points, while discrimination against daughters raises their mortality by 37 per thousand points. We replicate this analysis in the neonatal and postneonatal periods to take into account the fact that factors contributing to male-female differential mortality rates might differ across ages, but our results are qualitatively similar.

---

[13]See, e.g., Sen (1990b, 1992), Lin and Luoh (2008), Behrman (1988), Basu (1989), Oster (2009), and Ebeinstein (2006) for South and East Asia; and Sen (1990b), Sen (1992), Garenne (2003), and Deaton (2001) for sub-Saharan Africa.

[14]In studies comparing sub-Saharan Africa to India or other countries of South and East Asia, Sen (1990b, 1992) and Oster (2009) implicitly make a similar assumption.

We now apply our decomposition to child mortality. We find that the male-female difference in child mortality rates estimated from the sample of all twins is 4 and -16 per thousand points in sub-Saharan Africa and India, respectively. The male-female difference in infant mortality rates estimated from the sample of male-female twin pairs is respectively -8 and -31 per thousand points in these regions. These results imply that prenatal environmental factors raise the mortality rate of boys by 12 and 15 per thousand points in sub-Saharan Africa and India, respectively; the biological make-up of boys lowers their mortality rate by 8 per thousand points in both regions. Parental discrimination against daughters in India increases their mortality rate by 23 per thousand points.

The findings of this study lead to at least three important conclusions: (1) unobserved prenatal environmental factors account for a large fraction of excess male mortality observed in most populations; (2) the biological make-up of male children contributes to their excess mortality during infancy only, but its effect has been overestimated by about 50 percent in previous studies due to failure to account for prenatal factors;[16] and (3) parental discrimination against females in India increases females' mortality rates, although conventional methodological approaches underestimate the effects of discrimination by about 237 percent during infancy and 44 percent during childhood.

---

[15]The male-female difference in mortality being -10 per thousand points in India simply means that the mortality rate is on average 10 per thousand points lower for a male twin compared to his female co-twin.

[16]Contrary to the long-held biological theory explaining the sex gap in morbidity and mortality, the biological make-up of male children favors their survival during childhood.

## 1.2 Related literature

Comparing South and East Asia to sub-Saharan Africa is not new in the literature on gender discrimination. Amartya Sen (1990, 1992, 2003) uses sub-Saharan Africa, which has a population sex ratio close to the natural or biological sex ratio, as a benchmark to estimate the number of women that are "missing" in some countries of South and East Asia and North Africa due to discrimination against them.[17]

In a more recent study, Oster (2009) estimates the effect of discrimination on excess female mortality in India. The empirical strategy in this study accounts for the natural or biological survival advantage that female children have in societies that do not discriminate against them. In the analysis, such societies are assumed to be sub-Saharan Africa. Also, the effect of sex differences in biology is explicitly assumed to be the same in India and sub-Saharan Africa. Oster finds that discrimination leads to higher female mortality only after the first six months of life.[18]

By comparing sub-Saharan Africa with India, our paper shares similar motivations with prior work by Sen (1990, 1992, 2003) and Oster (2009), but it significantly differs in its scope, methodology, and results. To the best of our knowledge, our study is the first to decompose the sex gap in mortality into the effects of three important factors: prenatal environment, child biology, and female discrimination. A distinctive feature of our analysis is in explicitly taking into account the fact that offspring sex ratio is not random. In doing so, we find that prenatal environment contributes to excess male mortality, which is a new contribution to the literature.

We also find males' biology to contribute to their mortality during infancy, although its effect has been overestimated in previous studies. Nevertheless, our finding is consistent with a very large literature on biology that has documented sex differences in immune systems, neurodevelopmental disorders, genetic disorders, unintentional injuries, lendocrine response to perinatal stress, and tolerance of prenatal and postnatal malnutrition (e.g., Waldron (1983, 1985, 1998), Chao (1996)). Also, our finding that biology favors male survival during childhood may be consistent with the fact that females suffer a higher incidence of autoimmune diseases compared to males at certain ages, despite having a stronger immune system (Ansar Ahmed et al. (1985), Chao (1996), Bouman et al. (2005)).[19]

---

[17] As noted by Sen, comparing South and East Asia to sub-Saharan Africa is appropriate because these regions have similar mortality profiles. For the same reason, Western countries, which are more developed, cannot serve as a useful counterfactual.

[18] Oster (2009) pools data from India and selected sub-Saharan African countries, and estimates a linear probability regression of mortality on gender, a dummy indicator for India, and an interaction term between gender and India. For mortality occurring within six months after birth, the coefficient on the interaction term is not statistically different from zero, which means that the effect of discrimination on female newborns is negligible. Oster's methodology, however, improves over the conventional approach.

Further, we find that discrimination leads to substantially higher female mortality, even in the newborn period, than has been estimated in previous studies or in analysis of twin samples without controlling for twin fixed effects. Numerous studies have documented the effect of parental bias against daughters during early childhood (that is, between 12-60 months). However, by not controlling for prenatal and biological factors, most of them have failed to detect any such effect during the newborn period.[20] This has led some to claim that female discrimination has damaging effects only during childhood (e.g., Sen (1999), Osmani and Sen (2003)), which is inconsistent with the long documented existence of infanticide, neglect and abandonment of female newborns in South and East Asia (e.g., Sudha and Rajan (1999), Coale and Banister (1994)). Our decomposition methodology yields results that are consistent with this fact.

Finally, we view our study as a contribution to the few papers by economists who have used twin samples to study later life outcomes, while at the same time addressing a number of important methodological issues (e.g., Almond, Chay, and Lee (2005), Royer (2009), Behrman and Rosenweig (2004), Oreopoulos et al. (2006), Black, Devereux, and Salvanes (2007)). While most of these papers analyze data collected in developed countries, our study is among the very few based on large samples of twins from developing countries.

## 1.3 Plan of the paper

The remainder of this paper is organized as follows. Section 2 summarizes the literature on how prenatal environment affects offspring sex ratio. Section 3 develops the methodology. In Section 4, we describe our data, and show our results in Section 5. Section 6 discusses the effects of intrauterine environment as well as biology-environment interactions. Section 7 summarizes the key findings and concludes our study.

---

[19] Analyzing national data from the World Health Organization, Garenne (1992) finds that mortality from measles is higher for females compared to males. This is consistent with Preston (1976) who finds excess female mortality from certain diseases including for instance tuberculosis at ages 5-29, influenza-pneumonia-bronchitis at ages 5-14, and certain infectious and parasitic diseases at ages 1-14.

[20] As mentioned earlier, infant mortality is 27 per thousand points higher for male twins than female twins in India. Not accounting for the effects of prenatal environment and biology would therefore lead one to conclude to an absence of discrimination against daughters during infancy. However, controlling for those factors shows that discrimination substantially increases female infant mortality by 37 per thousand points.

## 2 Prenatal environment and offspring sex ratio

Recent studies show that offspring sex ratio is related to parental circumstances and levels of hormones at the time of conception. Levels of parental hormones are in turn related to parental stresses, illnesses and occupations (James (1994, 1995, 1996, 1997, 1998, 2001)). James (1998) provides evidence that men with multiple sclerosis or non-Hodgkins lymphoma are more likely to bear female children. Similarly, men engaged in professional driving or professional diving, and those exposed to the nematocide DBCP, dioxin, borates, vinclozolin, or high voltage installations bear excesses of daughters. However, men suffering from prostate cancer or treated with gonadotrophin or methyltestosterone are more likely to produce sons.

The work of James is consistent with many other studies on the effects of parental exposure to environmental toxicants such as 2,3,7,8-Tetrachlorodibenzo-p-dioxin (TCDD), fungicide, trichlorophenate, alcohol, lead, solvents, waste anesthetic gases and air pollution from incinerators on sex ratio.[21]

Works by a few economists also show that offspring sex ratio is not randomly determined. Almond and Edlund (2007) find a correlation between social class and sex ratio at birth in the United States. Almond and Mazumber (2009) find that maternal fasting during Ramadan affects sex ratio at birth, and Norberg (2004) finds a strong association between parental household composition at the time of a child's conception and sex ratio.

The effects of some prenatal factors on sex ratio are large enough to constitute a source of bias in studies examining the determinants of sex differences in child outcomes after birth. The proximal mechanisms are not well understood, but a variety of possibilities have been proposed. There is no testing of proximal mechanisms in this study.

In the next section, we show how our methodological approach overcomes the potential bias introduced by prenatal environmental correlates of child sex when estimating the determinants of sex differences in mortality. More precisely, we show how sex differences in mortality can be decomposed into the distinct effects of prenatal environment, child biology and discrimination.

---

[21]See, e.g., Mocarelli et al. (1996), Mocarelli et al. (2000), Moller (1998), Jacobsen et al. (2000), Gary et al. (1996), Gary et al. (2002), Dimid-Ward et al. (1996), James (1997), Williams et al. (1992), and Davis et al. (1998)

# 3 Econometric model

## 3.1 Estimating sex differences in mortality

The sex gap in mortality is usually estimated using the following specification:

$$M_{iht} = \theta Male_i + X_{ht}\pi + \in_{iht} \quad (1)$$

where $M_{iht}$ is a dummy variable indicating whether child $i$, born to parents $h$, died at time $t$ ($M_{iht}$ takes on value 1 if $i$ died at time $t$ and 0 if not); $Male_i$ is a dummy indicator for whether child $i$ is male; $X_{ht}$ is a vector of observed parental and household characteristics thought to be correlated with sex and mortality[22]; and $\in_{iht}$ is an error term usually assumed to be uncorrelated with sex. The parameter of interest $\theta$, which measures male-female difference in mortality rates, is generally interpreted as the effect of inherent biological differences between boys and girls, and/or the effect of parental discrimination against females. Its interpretation, however, is very ambiguous in the literature. When $\theta$ is positive (meaning that the probability of dying is greater for boys than girls), this is generally interpreted as the effect of boys' weaker biology; and when $\theta$ is negative, this is interpreted as the effect of parental discrimination against girls. As discussed in the introduction, the observation that *infant mortality* is higher for boys than girls in both sub-Saharan Africa and India (Figure 1) has led many studies to claim that parental discrimination has a negligible effect on girls' mortality during infancy in India. We will show that such a claim mainly lies in the fact that sex (that is, $Male_i$ in equation (1)) is treated as exogenous in those studies.

The assumption made in most studies that $\in_{iht}$ is uncorrelated with child sex is not plausible in view of the evidence provided in Section 2, which shows that the sex of a child is determined by prenatal factors that might also affect health and survival. Any estimate of $\theta$ that does not address this issue of omitted variable bias is therefore likely to be misinterpreted and biased, although the direction of the bias is difficult to determine.[23] Our goal in this study is to overcome this bias. More precisely, we decompose $\theta$ into the effects of prenatal environmental factors, child biology and parental preferences.

---

[22] Note however that since sex has been traditionally treated as exogenous, controlling for the vector $X_{ht}$ is irrelevant in most studies.

Write $\epsilon_{iht} = u_h + v_{ht} + w_{iht}$ where $u_h$, $v_{ht}$ and $w_{iht}$ are respectively parental time-invariant unobservables, parental time-variant unobservables, and a child random unobserved shock (not correlated with sex).[24] $u_h$ and $v_{ht}$ are interpreted as parental prenatal circumstances and gender bias, respectively. We can explicitly write $v_{ht}$ as the sum of time-variant parental prenatal circumstances $(p_{ht})$[25] and parental bias $(b_{ht})$; that is, $v_{ht} = p_{ht} + b_{ht}$. They are correlated with child sex and mortality. We posit that a cross-sectional linear probability model (LPM) estimate of $\theta$ is the additive effects of child biology, prenatal factors, and parental preferences.[26] The effect of parental time-invariant unobservables can be netted out by comparing the mortality of male-female siblings (children born to the same parents). This is done by estimating a family fixed effect regression as follows:

Let $(i, j)$ be a pair of siblings. Re-writing Equation (1) for $i$ and $j$ yields respectively Equations (2) and (2)' below.

$$M_{iht} = \theta_{SFE} Male_i + X_{ht}\pi + u_h + p_{ht} + b_{ht} + w_{iht} \quad (2)$$
$$M_{jht'} = \theta_{SFE} Male_j + X_{ht'}\pi + u_h + p_{ht'} + b_{ht'} + w_{jht'} \quad (2)'$$

Taking (2)-(2)' yields:

$$M_{iht} - M_{jht} = \theta_{SFE}(Male_i - Male_j) + (X_{ht} - X_{ht'})\pi + p_{ht} - p_{ht'} + b_{ht} - b_{ht'} + w_{iht} - w_{jht'} \quad (3)$$

Estimating Equation (3) using a family fixed effect regression yields an estimate of $\theta_{SFE}$. Note that $\theta_{SFE}$ still includes the effect of parental time-variant factors as long as $p_{ht} - p_{ht'}$, for instance, is correlated with child sex (parental environmental and health circumstances, for instance, are likely to vary over time, making such a correlation very likely.) To net out the effect of prenatal factors, we compare a male twin with his female co-twin, by estimating a twin fixed effect regression. Let $(i, -i)$ be a pair of male-female twins. Equation (1) can be re-written for each of them as follows:

$$M_{iht} = Male_i \theta_{TFE} + X_{ht}\pi + u_h + p_{ht} + b_{ht} + w_{iht} \quad (4)$$
$$M_{-iht'} = Male_{-i} \theta_{TFE} + X_{ht}\pi + u_h + p_{ht'} + b_{ht'} + w_{-iht'} \quad (4)'$$

---

[23] Exposures to particular diseases or treatments, for instance, may lead not only to excess male births, but may also contribute to male mortality. If this is the case, then the share of excess male mortality generally attributed to biology is exaggerated. But if boys are also more likely to be born to parents of higher socioeconomic status (Almond and Edlun (2007)), which also contributes to child survival, then the share of excess male mortality attributed to biology is underestimated.

[24] This additive model follows from biological models often used to disentangle the effects of genetic and environmental factors on health outcomes (e.g., Evans et al. 2002, Neale and Cardon 1992). Fowler et al. (2008) also apply an additive model to a sample of twins to study the impact of unobserved genetic factors on political participation.

[25] Parental prenatal circumstances determining offspring sex ratios such as occupation or exposure to dioxin might vary over time.

[26] Note that if $u_h$ and $v_{ht}$ were observed and controlled in equation (1), $\theta$ would only measure the effect of male biology.

Since prenatal (or more precisely preconception) factors are the same for a pair of twins (that is, $p_{ht} = p_{ht'}$), taking (4)-(4)' yields:

$$M_{iht} - M_{-iht} = \theta_{TFE}(Male_i - Male_{-i}) + b_{ht} - b_{ht'} + w_{iht} - w_{-iht'} \quad (5)$$

$b_{ht} - b_{ht'}$ is still correlated with $Male_i - Male_{-i}$, which implies that estimating equation (5) using within male-female twin fixed effect regression yields an estimate of $\theta_{TFE}$, which is the additive effects of child biology and parental bias. Note that subtracting $\theta_{TFE}$ from the cross-sectional LPM estimate of $\theta$ yields an estimate of the effect of prenatal factors. Also, in societies where parents do not discriminate against any specific sex in the allocation of household resources (that is, $b_{ht} - b_{ht'} = 0$), $\theta_{TFE}$ measures the effect of male biology since $w_{iht} - w_{-iht}$ is uncorrelated with $Male_i - Male_{-i}$ by assumption.

We estimate $\theta$ and $\theta_{TFE}$ using samples of twins, which allows us to separate out the effects of prenatal factors, child biology and parental preferences (see Section 3.2 below).

## 3.2 Decomposing sex differences in mortality into the effects of prenatal factors, child biology, and parental preferences

We posit that $\theta$, which is the male-female difference in mortality or the effect of sex on mortality estimated from Equation (1), is the additive effects of prenatal factors, child biology and parental preferences.[27] That is:

---

[27] As noted earlier, this simplifying additivity assumption largely follows from biological models often used to disentangle the effects of genetic and environmental factors on health outcomes (e.g., Evans et al. 2002, Neale and Cardon 1992). In Section 6, we discuss some results supporting this assumption.

$$\theta = \theta_1 + \theta_2 + \theta_3 \quad (6)$$

where $\theta_1$ is the effect of prenatal factors, $\theta_2$ the effect of child biology, and $\theta_3$ the effect of parental preferences.

We assume that parental discrimination varies from one society to another. More precisely, we assume two types of societies: non-discriminatory (ND) and discriminatory (D). The effect of parental discrimination in a non-discriminatory society is zero by definition. We also assume that the effect of male biology on sex differences in mortality is the same in discriminatory and non-discriminatory societies. Both assumptions can be formally expressed as follows.

$$\begin{cases} \theta_3^{ND} = 0 \\ \theta_2^{D} = \theta_2^{ND} \end{cases} \quad (7)$$

Plugging the first and the second equation of System (7) into Equation (6), and re-writting Equation (6) for discriminatory and non-discriminatory societies, respectively, yields:

$$\begin{cases} \theta^{ND} = \theta_1^{ND} + \theta_2^{ND} \\ \theta^{D} = \theta_1^{D} + \theta_2^{D} + \theta_3^{D} \end{cases} \quad (8)$$

We then separate out $\theta_1^{ND}$ and $\theta_2^{ND}$ on one hand, and $\theta_1^{D}$, $\theta_2^{D}$ and $\theta_3^{D}$ on the other hand. Estimating Equation (1) yields $\theta^{ND}$ and $\theta^{D}$ for non-discriminatory and discriminatory societies, respectively. Estimating a twin fixed effect regression (Equation (5)) yields $\theta_{TFE}$, which is the additive effects of biology and parental preferences. Given that the effect of parental preferences is zero in non-discriminatory societies, $\theta_{TFE}$ in these environments solely measures the effect of biology. That is:

$$\begin{cases} \theta_{TFE}^{ND} = \theta_2^{ND} \\ \theta_{TFE}^{D} = \theta_2^{D} + \theta_3^{D} \end{cases} \quad (9)$$

Plugging the first and second equation of System (9) into the first and second equation of System (8), respectively, and re-arranging, allows us to extract the effect of prenatal factors as follows:

$$\begin{cases} \theta_1^{ND} = \theta^{ND} - \theta_{TFE}^{ND} \\ \theta_1^{D} = \theta^{D} - \theta_{TFE}^{D} \end{cases} \quad (10)$$

We have separated out the effects of prenatal factors and biology in non-discriminatory societies. For discriminatory societies, it remains to separate out the effects of child biology and parental preferences. Remember that the effect of biology is the same in non-discriminatory and discriminatory societies: $\theta_2^D = \theta_2^{ND} = \theta_{TFE}^{ND}$. Plugging this latter equation into the second equation of System (9) and re-arranging yields the effect of parental preferences:

$$\theta_3^D = \theta_{TFE}^D - \theta_{TFE}^{ND} \quad (11)$$

We summarize the results obtained from System (7) through Equation (11) in System (12) below:

$$\begin{cases} \theta_1^{ND} = \theta^{ND} - \theta_{TFE}^{ND} \\ \theta_2^{ND} = \theta_{TFE}^{ND} \\ \theta_3^{ND} = 0 \\ \theta_1^{D} = \theta^{D} - \theta_{TFE}^{D} \\ \theta_2^{D} = \theta_{TFE}^{ND} \\ \theta_3^{D} = \theta_{TFE}^{D} - \theta_{TFE}^{ND} \end{cases} \quad (12)$$

System (12) separates out the roles of prenatal factors, child biology and parental preferences in sex differences in mortality in discriminatory and non-discriminatory societies.

## 3.3 Empirical strategy

### 3.3.1 Estimating the effects of prenatal factors, child biology, and parental preferences across ages

It is possible that the effects of prenatal factors, child biology or parental preferences on sex differences in mortality vary with age. We distinguish two main periods of child development:

infancy (I) and childhood (CH). Infant mortality is measured as the probability of dying during the first year of life. Child mortality is measured as the probability of dying between the first and fifth birthdays, conditional on surviving the infant period.

In most developing countries, mortality occurring during the newborn period accounts for a large fraction of under-five mortality (Black et al. (2003)). Therefore, we further divide up the infant period into the neonatal (NN) period (that is, the period from birth to 28 days or 1 month), and the postneonatal (PNN) period (1-12 months). Neonatal mortality is measured as the probability of dying during the neonatal period conditional on being born alive, and postneonatal mortality is the probability of dying during the postneonatal period, conditional on surviving the neonatal period.

We estimate Equations (1) and (5) for each of the periods just defined. When estimating the twin fixed effect regression (Equation (5)) during the postneonatal or childhood period, we drop out surviving twins whose counterparts did not survive the preceding period. Our decomposition of the sex gap leads to the derivation of parameters in Equation (12) for each period (P) as follows:

$$\begin{cases} \theta_{1,P}^{ND} = \theta_P^{ND} - \theta_{TFE,P}^{ND} \\ \theta_{2,P}^{ND} = \theta_{TFE,P}^{ND} \\ \theta_{3,P}^{ND} = 0 \\ \theta_{1,P}^{D} = \theta_P^{D} - \theta_{TFE,P}^{D} \\ \theta_{2,P}^{D} = \theta_{TFE,P}^{ND} \\ \theta_{3,P}^{D} = \theta_{TFE,P}^{D} - \theta_{TFE,P}^{ND} \end{cases} \quad \text{with } P = I,\ NN,\ PNN,\ \text{or } CH \quad (13)$$

### 3.3.2 Choice of discriminatory and non-discriminatory societies

Our choice of discriminatory and non-discriminatory societies is based on studies conducted in different regions. It is well documented that in most South and East Asian countries, parents have strong pro-male bias, discriminating against female children in the allocation of food, health care, and other resources (e.g., Sen1990b, 1992, 2003)). On the contrary, there is little evidence of such discrimination in sub-Saharan Africa. Based on household data, Garenne (2003) finds that the probability of dying before the fifth birthday is higher for boys than girls, but investment in health care such as immunization does not significantly differ between the two sexes. Further evidence for the symmetrical treatment of boys and girls in a sub-Saharan African country is provided by Deaton (2001). Using household expenditure data from Côte d'Ivoire, Deaton (2001) finds no gender bias in the allocation of goods, while finding a statistically non-significant pro-male bias in Thailand. The findings of these studies support the assumption that sub-Saharan Africa is non-discriminatory, as also recognized by Sen (1990b, 1992, 2003).

For our analysis, we use data from India, considered a discriminatory society, and from sub-Saharan African countries, considered non-discriminatory. Our assumption that biology has similar effects in these regions is consistent with Sen (1990, 1992, 2003) and Oster (2009). The plausibility of this assumption also relies on the fact that children in these regions suffer and/or die from similar diseases (Black et al. (2003)).

# 4 Data

## 4.1 Data sources

We use Demographic and Health Surveys data collected in thirty sub-Saharan African countries, and two National Family Health Surveys conducted in India. Information on years of surveys is provided in Table A-1 in the appendix. The DHS and the NFHS surveys are conducted by the same organization (Measure DHS), and are standardized and comparable across countries and years for most variables. In each survey, a two-stage probabilistic sampling technique is used to select clusters or census enumeration zones at the first stage, and household at the second stage. In each household, data are collected on characteristics including durable assets and facilities (e.g., car, TV, radio, access to clean water, toilet facilities).

Information on the demographic and socioeconomic characteristics of each household member is also collected. Selected women in the household provide complete information on their fertility history. In particular, information is provided on each live birth, including date of birth, whether the birth is a singleton or a multiple birth, whether the person is still alive or not, and when the person died if dead. In this study, we use the file of all live births reported by mothers in each survey. The number of these files is 75 for sub-Saharan Africa, and 2 for India. Sub-Saharan African countries' files are merged and analysed as a single file, and so are the two files from India. The total sample size of all live births is 1,670,477 for sub-Saharan Africa, and 543,981 for India. Detailed information on the sample size of all births for each country and survey year can be found in Table A-1.

## 4.2 Data descriptions

In describing the data, we show how twins and singletons are comparable along several demographic and socioeconomic factors. The goal of this comparison is to show that twins are not a selected population with respect to those factors. While this comparison provides useful information, our goal is not to generalize results obtained from analyzing twin samples to the entire population.

### 4.2.1 Sex ratios

Information on whether a birth was single or multiple is provided in each survey. We identified and matched twins based on: (1) whether they were declared as twins by their mothers; (2) their mother's identification number; and (3) their month and year of birth. Triplets and quadruplets are dropped from the sample. Table 1 shows that the sample size of twins is 50,994 for sub-Saharan Africa and 6,920 for India. They represent respectively 3.05 percent and 1.27 percent of the sample of all live births in these settings. Twinning rates vary across sub-Saharan Africa (Table A-1), but the reasons are not entirely known. The proportion of twins for sub-Saharan Africa is comparable to that found in the United States by Almond, Chay and Lee (2005). This proportion seems to be low for India, but the reasons for this are not known. In sub-Saharan Africa, male-male, female-female, and male-female twins represent respectively 31%, 30% and 39% of all twins. In India, these figures are respectively 35%, 33% and 32%.

We note that the proportion of male births is 0.508 and 0.504 for singletons and twins, respectively, in sub-Saharan Africa, and 0.520 and 0.514 in India. This indicates a slightly lower proportion of male among twins in both settings. However, the relevant comparison of sex ratios should be between singletons and same-sex twins. The proportion of males among same-sex twins is 0.506 and 0.521 in sub-Saharan Africa and India, respectively, figures that are similar to the proportion of males among singletons in these regions (0.508 and 0.520, respectively). This suggests that male-female relative differences in fetal death are consistent with those for twins and singletons, and so are the prenatal environmental factors that determine the sex of a child.

For the pooled sample of twins and singletons, these figures imply a sex ratio at birth (the ratio of males to females at birth) of 1.032 in sub-Saharan Africa and 1.08 in India. The figure for sub-Saharan Africa is similar to that found by Garenne (2002) based on Demographic and Health Surveys and World Fertility Surveys. The figure for India is the same as that found both in the 2001 Indian Census, as well as earlier work by Rosenzweig and Schultz (1982) that was based on a nationally representative sample of rural households in India. This figure is also consistent with the sex ratio of 1.09 found by Pakrasi and Halder (1971) using the 1961-62 Indian National Sample Survey. High sex ratios at birth in India have been explained by the selective abortion of female fetuses (e.g, Sen (1990b, 1992), Ebenstein (2007)). Sex ratios in sub-Saharan Africa and India are significantly different from the world sex ratio of 1.055, but are closer to what some biologists have called the "natural sex ratio."

### 4.2.2 Socioeconomics

In Table 2, we show the summary statistics of common demographic and socioeconomic variables for twins and singletons. In sub-Saharan Africa and India, twins and singletons are similar along several characteristics such as maternal age, marital status and education, and paternal education. Twins tend to be born to slightly older mothers than singletons in both regions. In India, a slightly higher proportion of twins than singletons are born to mothers or fathers with a secondary or higher level education. With respect to household characteristics, we note that twins obviously live in slightly larger households than singletons, but they do not significantly differ along other household characteristics including wealth (e.g., electricity, radio, TV, car). This comparison show that twins largely mirror the entire population along several demographic and socioeconomic variables, as found in other studies (e.g., Almond, Chay and Lee (2005)).

### 4.2.3 Mortality

Mortality rates are much higher for twins than singletons (Table 2). In sub-Saharan Africa, the probability of dying before the fifth birthday is 163 per thousand for singletons and 405 per thousand for twins. These figures are respectively 115 and 329 per thousand in India. Differences in mortality rates decline with age. The twin-singleton mortality rate ratio in the neonatal period is close to 5 in sub-Saharan Africa (193 vs. 41 per thousand), and 6 in India (287 vs. 50 per thousand). This ratio falls below 2 in both regions by age 5.

The comparison between twins and singletons shows that while twins are not selected with respect to common demographic and socioeconomic factors, being a twin has a positive effect on mortality.

#### 4.2.4 Generalizability

It is also important to note that while it is theoretically appealing to estimate the effects of prenatal environment, biology and parental discrimination on sex differences in mortality using twins, like in most twin studies, there are potential threats to external validity. Although twins and singletons are comparable along several socioeconomic dimensions, their mortality differs substantially. It is therefore not clear how our estimates obtained from twin samples generalize to the population of singletons. However, because offspring sex ratio is similar for same-sex twins and singletons, this suggests that prenatal factors that determine child sex are similar for same-sex twins and singletons.[28] Our analysis therefore suggests that prenatal factors are important in determining sex differences among singletons as well, although we do not know the magnitude of the effect. It is also worth noting that the focus on twins is useful, as twins constitute an important and fast growing population that needs to be studied.[29]

## 5 Results

### 5.1 Sex differences in mortality: Descriptive analysis

---

[28] Among mixed-sex twins, it is obvious that child sex is perfectly uncorrelated with prenatal environment.

[29] In 2006, 125 million of the world population were twins (Oliver (2006)). The population of twins is growing very fast due to assisted reproduction, especially in developed countries.

In Table 3, we show the mortality rates of males and females at different ages for all twins and for male-female twins in sub-Saharan Africa and India. For all twins, infant mortality is higher among males than among females (323 vs. 277 per thousand in sub-Saharan Africa, and 393 vs. 366 per thousand in India). The sex gap in infant mortality rate is 46 per thousand points in sub-Saharan Africa and 27 per thousand points in India. In the sample of male-female twins, the sex gap drops to 27 per thousand points in sub-Saharan Africa, and completely reverses in India, where infant mortality is now 10 per thousand points higher among girls than boys. Given that male-female twins have the same exposure to prenatal factors (here defined as factors occurring before conception), the smaller sex gap found in the sample of male-female twins in sub-Saharan Africa clearly rules out the effect of these factors, and the reversed gap in India additionally shows the effect of discrimination against female children.

In the neonatal, postneonatal and childhood periods, the results are qualitatively the same as in the infant period. We however note that while female children have a survival advantage in the neonatal period, they die at a higher rate in subsequent periods in India, while still keeping their advantage in sub-Saharan Africa.

## 5.2 Decomposing sex differences in mortality into the effects of prenatal factors, child biology and parental preferences

### 5.2.1 Infant mortality

We estimate sex differences in mortality for "all twins" (equation (1)) and for "mixed-sex twins" (this is equivalent to estimating a twin fixed effect LPM: equation (5)).[30] The dependent variable is a dummy indicator taking on the value 1 if the child died before his/her first birthday, and 0 if not. Results are presented in Panel A of Table 4. Columns (I)-(III) show the results for sub-Saharan Africa, and Columns (IV)-(VI) show the results for India. In Columns (I) and (IV), the dependent variable is regressed on a binary variable taking on the value 1 if the child is male, and 0 if the child is female. As shown in the descriptive analysis, the probability of dying before the first birthday is 47 and 27 per thousand points higher among males than females in sub-Saharan Africa and India, respectively. In Columns (II) and (V), we control for child, parental and household characteristics.[31] This changes little from the estimates obtained in Column (I) and (IV). The male-female difference in infant mortality decreases to 45 per thousand points in sub-Saharan Africa, but remains the same in India.

---

[30]Note that we use a linear probability model to facilitate the exposition of the results. Using a probit model changes little to our findings.

The existence of unobserved prenatal factors that may affect both child sex and health, as suggested by the biological literature surveyed in section 2, implies that the estimates of the sex gap in infant mortality obtained in Columns (I)-(II) and (IV)-(V) are biased. We correct for this bias by estimating a twin fixed effect regression in Column (III) for sub-Saharan Africa and Column (VI) for India. Infant mortality is now only 27 per thousand points higher for boys compared to girls in sub-Saharan Africa, and is 10 per thousand points lower in boys compared to girls in India. However the estimate for India is not statistically significant at the 10% level.

In Table 5, we show the decomposition of sex differences in infant mortality into the effects of prenatal environment, child biology and parental preferences. These estimates are computed based on the point estimates obtained in Columns (II) and (III) for sub-Saharan Africa, and Columns (V) and (VI) for India. From this calculation, we find that prenatal environmental factors play a significant role in the differential mortality rates of male and female children. These factors raise boys' infant mortality rate by 45-27=18 per thousand points in sub-Saharan Africa and by 27 -(-10)=37 per thousand points in India. Biology is also an important factor in the sex gap in mortality, but its role is much less important than previously thought. Males' biology increases their mortality rate by 27 per thousand points in sub-Saharan Africa and India.

Finally, discrimination against female children in India increases their mortality rate by 37 per thousand points. This finding contradicts previous conclusions that discrimination against female *infants* in South and East Asia had a negligible effect on their mortality. As noted earlier, such conclusions are based on the fact that infant mortality is higher among boys than girls, as shown in Figure 1 for India. Our analysis shows that once prenatal environment and biology are controlled for, the effect of discrimination against girls becomes more visible, a finding which is consistent with well-documented evidence of neglect, aban-donment, and infanticide inflicted on female newborns in most countries of South and East Asia (Sudha and Rajan (1999)).

---

[31]These characteristics include child's year of birth, mother's age at survey, education and marital status, husband's education, household size, possession of household assets and facilities, a linear control for year of survey, and country fixed effect.

### 5.2.2 Neonatal mortality

We repeat the analysis for neonatal mortality. The results are presented in Table 4, Panel B. Columns (I) and (IV) show that neonatal mortality is 37 and 43 per thousand points higher among male children than female children in sub-Saharan Africa and India, respectively. After controlling for child, parental and household characteristics in Columns (II) and (V), the coefficient on the male dummy decreases to 36 per thousand points for sub-Saharan Africa, and increases to 45 per thousand points for India. We estimate a twin fixed effect regression in Columns (III) and (VI). We find that neonatal mortality is now only 22 and 9 per thousand points higher among males than females in sub-Saharan Africa and India, respectively, but the estimate for India is not statistically different from zero.

In Table 5, we show the decomposition of sex differences in neonatal mortality. Prenatal environmental factors increase the neonatal mortality rate of male children by respectively 14 and 36 per thousand points in sub-Saharan Africa and India, respectively. The biological make-up of male children raises their mortality by 22 per thousand points, and parental discrimination against female children in India increases their mortality by 13 per thousand points.

### 5.2.3 Postneonatal mortality

The results for postneonatal mortality are shown in Table 4, Panel C. Columns (I) and (IV) show that postneonatal mortality is 18 per thousand points higher for boys than girls in sub-Saharan Africa, but is 14 per thousand points lower for boys in India, although the estimate for India is not statistically significant at the level 10%. Adding controls changes little to those estimates (Columns (II) and (V)). We estimate a twin fixed effect regression in Columns (III) and (VI). Postneonatal mortality is now only 10 per thousand points higher among males than females in sub-Saharan Africa, and is 32 per thousand points higher among females than males in India.

The results of the decomposition analysis are presented in Table 5. We note that prenatal environment increases the mortality of male children by 8 and 18 per thousand points in sub-Saharan Africa and India, respectively. The biological make-up of male children contributes 10 per thousand points to their mortality rate, and parental discrimination against female children in India raises their mortality by 42 per thousand points.

### 5.2.4 Child mortality

The results for child mortality are presented in Table 4, Panel D. Columns (I) and (IV) show that child mortality is 4 per thousand points higher among males in sub-Saharan Africa (results not statistically significant), but is 17 per thousand points lower in this sex in India. Adding controls changes little to these estimates (Columns (II) and (V)). In Columns (III) and (VI), we estimate a twin fixed effect regression. Child mortality is now 8 and 31 per thousand points lower among male children compared to female children in sub-Saharan Africa and India, respectively.

In Table 5, we show the decomposition of the sex gap in mortality. We note that prenatal environment raises the mortality of male children by 12 and 15 per thousand points in sub-Saharan Africa and India, respectively. But contrary to the long-held biological theory of sex differences in morbidity and mortality, the biological make-up of male children now favors their survival during the childhood period. Biology reduces male mortality by 8 per thousand points. Parental discrimination against female children in India increases their mortality by 23 per thousand points.

# 6 Discussions: Zygosity, intrauterine environment, and biology-environment interactions

Our twin samples include both identical and fraternal twins. However, like in other twin studies (e.g., Almond, Chay, and Lee (2005), Royer (2009), Oreopoulos et al. (2006)), our data do not allow us to distinguish between both types in our analysis, which means that we can only estimate the effect of *preconception* environment, not *intrauterine* environment, on sex differences in mortality. Our estimates rely on comparing "all twins" with "male-female twins." Among all twins, same-sex twins are often, but not always identical, while male-female twins are always fraternal. Identical twins often have perinatal problems (due, for instance, to sharing the same placenta) that fraternal twins do not have, which often results in lower birth weight in the former. Therefore, our estimates of the effect of preconception environment would extend to intrauterine environment only if we assume that perinatal problems due to monozygosity have a similar mortality effect on male twin pairs and female twin pairs. Such an assumption would perhaps be consistent with the literature that suggests that zygosity may not be so critical for estimating the effect of intrauterine environment on later life outcomes.[32] In our study, we only focus attention on preconception environment, which interestingly precedes intrauterine environment, and which, due to the fact that it determines child sex, can be viewed as a distal determinant of monozygosity among same-sex twins.[33] It would therefore be interesting in a future research to determine how much of the effect of preconception environment is mediated by monozygosity.[34]

Also, while we do not have good data on birth weight, a usual proxy for intrauterine growth, we control for birth order within twin pairs in results not shown. It has been shown that twin first-borns are heavier and have lower mortality rates than twin second-borns (e.g., Smith, Pell and Dobbie (2002), Buekens and Wilcox (1993)), suggesting that birth order is a proxy for intrauterine environmental conditions.[35] This has little effect on our results.

Further, following several studies of twins (e.g., Evans et al. (2002), Neale and Cardon (1992), Fowler et al. (2008)), we have assumed the effects of prenatal environment, biology and discrimination to be additive, abstracting away from potential interactions between those factors. In results not shown, we estimate the effect of child sex interacted with several proxies for environments. The interaction between sex and maternal education has no effect on mortality in sub-Saharan Africa or in India.[36] Similarly, interactions between sex and birth order (among all children and within twin pairs), or sex and climate zones (within Africa) have no effect on mortality at most ages. While the availability of richer information on prenatal environment would have allowed us to fully assess the additivity assumption made in this study, we also acknowledge that these results tend to support such an assumption.

---

[32]The medical literature suggests that adult health outcomes among identical twins are similar to those among fraternal twins (e.g., Christensen et al. (1995), Duffy (1993)). Also, Black, Devereux, and Salvanes (2007) find that the effects of birth weight on outcomes such as education and earnings are similar for identical and fraternal twins. Although the focus of these studies is not on examining sex differences in outcomes, they do however suggest that zygosity may not be so important for estimating the effects of intrauterine environment on later life outcomes.

[33]As noted earlier, among mixed-sex twins, it is obvious that child sex is perfectly uncorrelated with preconception environment. This implies that our estimates of the effects of biology and parental gender bias on gender differences in mortality are likely to be unbiased.

[34]Note that monozygous twins account for only 30% of all twins (MacGillivray (1986)). Of all monozygous twins, only 60-70% share the same placenta, and only 1-2% share the same amniotic sac. Therefore, only around 20% of all twins share the same placenta and less than 1% share the same amniotic sac. Also, we are not aware of any evidence that the effects of monozygosity on child outcomes differ by sex. It is therefore possible that the mediating effect of monozygosity be small in reality, consistent with prior studies focusing on much later life outcomes (e.g., Christensen et al. (1995), Duffy (1993), Black, Devereux, and Salvanes (2007)).

[35]We do also observe in our data that twin first-borns have lower mortality rates than twin second-borns, confirming earlier studies.

# 7 Concluding remarks

We have decomposed sex differences in infant and child mortality into the distinct effects of prenatal environment, child biology, and parental discrimination against daughters. Our methodology accounts for the fact that offspring sex ratio is not random, but is partly determined by prenatal factors that may also affect child health and survival after birth. In doing so, we have overcome bias in previous estimates of the effect of female discrimination on mortality. More precisely, using large samples of twins from sub-Saharan Africa and India, we have found that: (1) unobserved factors in the prenatal environment account for a large fraction of excess male mortality; (2) biological make-up of male children positively affects their mortality during infancy only, but this effect is usually overestimated due to failure to control for prenatal factors; (3) in India, parental discrimination against female children has a sizeable effect on their mortality; however, failure to control for prenatal and biological effects leads conventional approaches to underestimate its effect by about 237 percent during infancy, and 44 percent during childhood.

Understanding the origins of sex imbalance in mortality is essential in designing policies that efficiently address this crucial issue. Sex differences in biology have long been advanced as the unique explanation for the excess mortality of male children in non-discriminatory societies, leaving the impression that little could be done to improve their survival. The demonstrated role of prenatal environmental factors in male mortality means that actions can be undertaken to address this issue. Our analysis also shows the role of discrimination in increasing the mortality of female children in India. That this effect has been underestimated in previous studies simply means that new efforts should be undertaken by researchers, governments and policymakers to combat this very crucial problem that unjustly prevents millions of women from life.

---

[36]Education is a proxy for social class, which has been found to be a determinant of sex ratio (e.g., Almond and Edlund (2007), James (1998)). Also, in most developing countries, non-educated mothers are more likely to be employed in the agricultural sector (if indeed they are employed), thereby increasing their exposure to certain fertilizers and chemicals likely to affect the sex ratio of their offspring.

Table 1: Sex ratios at birth of singletons and twins in sub-Saharan Africa and India

|  | Sub-Saharan Africa | | India | |
|---|---|---|---|---|
|  | Sample size | % boys (S.D) | Sample size | % boys (S.D) |
| Singletons | 1,619,483 | 0.508 (0.500) | 537,061 | 0.520 (0.500) |
| All twins | 50,994 | 0.504 (0.500) | 6,920 | 0.514 (0.500) |
| Male-female | 20,154 | 0.500 (0.500) | 2,232 | 0.500 (0.500) |
| Male-male | 15,610 | 1 (0) | 2,442 | 1 (0) |
| Female-female | 15230 | 0 (0) | 2,246 | 0 (0) |
| Same-sex twins | 30,840 | 0.506 (0.500) | 4,688 | 0.521 (0.500) |

**Note:** The sex ratios of same-sex twins and singletons are very similar in both sub-Saharan Africa (0.506 *vs.* 0.508) and India (0.521 *vs.* 0.520), suggesting that prenatal environmental factors determining the sex of a child are similar for same-sex twins and singletons. Among male-female twins, child sex is obviously perfectly uncorrelated with prenatal factors.

Table 2: Summary statistics

| | Sub-Saharan Africa | | | | | | India | | | | | |
|---|---|---|---|---|---|---|---|---|---|---|---|---|
| | Singletons | | | Twins | | | Singletons | | | Twins | | |
| **Variables** | N | Mean | S.D | N | Mean | S.D | N | Mean | S.D | N | Mean | S.D |
| Child is male | 1,619,483 | 0.508 | 0.500 | 50,994 | 0.504 | 0.500 | 537,061 | 0.520 | 0.500 | 6,920 | 0.514 | 0.500 |
| **Maternal characteristics** | | | | | | | | | | | | |
| Age | 1,619,483 | 35.104 | 8.062 | 50,994 | 36.343 | 7.521 | 537,061 | 34.772 | 8.040 | 6,920 | 35.225 | 7.804 |
| Marital status | | | | | | | | | | | | |
|   Single | 1,619,432 | 0.022 | 0.148 | 50,994 | 0.015 | 0.122 | 537,061 | 0.000 | 0.000 | 6,920 | 0.000 | 0.000 |
|   Married | 1,619,432 | 0.769 | 0.422 | 50,994 | 0.771 | 0.420 | 537,061 | 0.943 | 0.232 | 6,920 | 0.942 | 0.233 |
|   Widowed | 1,619,432 | 0.047 | 0.211 | 50,994 | 0.050 | 0.218 | 537,061 | 0.047 | 0.212 | 6,920 | 0.047 | 0.211 |
|   Living with a partner | 1,619,432 | 0.097 | 0.296 | 50,994 | 0.096 | 0.295 | 537,061 | 0.000 | 0.000 | 6,920 | 0.000 | 0.000 |
|   Not living with a partner | 1,619,432 | 0.034 | 0.181 | 50,994 | 0.037 | 0.188 | 537,061 | 0.008 | 0.090 | 6,920 | 0.008 | 0.090 |
|   Divorced or separated | 1,619,432 | 0.031 | 0.173 | 50,994 | 0.030 | 0.172 | 537,061 | 0.002 | 0.042 | 6,920 | 0.003 | 0.051 |
| Education | | | | | | | | | | | | |
|   Not Educated | 1,619,404 | 0.554 | 0.497 | 50,990 | 0.558 | 0.497 | 536,070 | 0.624 | 0.484 | 6,908 | 0.597 | 0.491 |
|   Primary | 1,619,404 | 0.335 | 0.472 | 50,990 | 0.335 | 0.472 | 536,070 | 0.171 | 0.376 | 6,908 | 0.184 | 0.387 |
|   Secondary or higher | 1,619,404 | 0.111 | 0.314 | 50,990 | 0.107 | 0.309 | 536070 | 0.206 | 0.404 | 6,908 | 0.219 | 0.414 |
| **Father's education** | | | | | | | | | | | | |
|   Not Educated | 1,548,881 | 0.579 | 0.494 | 49,038 | 0.580 | 0.493 | 536,465 | 0.623 | 0.485 | 6,906 | 0.597 | 0.491 |
|   Primary | 1,540,365 | 0.352 | 0.478 | 48,576 | 0.351 | 0.477 | 535,280 | 0.171 | 0.376 | 6,888 | 0.184 | 0.388 |
|   Secondary or higher | 1,512,371 | 0.119 | 0.323 | 47,740 | 0.114 | 0.318 | 535,115 | 0.206 | 0.404 | 6,886 | 0.220 | 0.414 |
| **Household characteristics** | | | | | | | | | | | | |
| Household size | 1,619,483 | 7.993 | 4.795 | 50,994 | 8.447 | 4.728 | 537,061 | 7.229 | 3.539 | 6,920 | 7.427 | 3.772 |
| Has electricity (0/1) | 1,519,492 | 0.170 | 0.376 | 47,648 | 0.167 | 0.373 | 537,061 | 0.596 | 0.491 | 6,920 | 0.610 | 0.488 |
| Has radio (0/1) | 1,584,591 | 0.551 | 0.497 | 49,820 | 0.556 | 0.497 | 536,867 | 0.415 | 0.493 | 6,920 | 0.427 | 0.495 |
| Has TV (0/1) | 1,532,985 | 0.126 | 0.332 | 47,972 | 0.122 | 0.327 | 536,918 | 0.304 | 0.460 | 6,920 | 0.309 | 0.462 |
| Has car (0/1) | 1,527,477 | 0.042 | 0.201 | 47,950 | 0.039 | 0.193 | 536,897 | 0.016 | 0.127 | 6,920 | 0.019 | 0.137 |
| **Child outcomes** | | | | | | | | | | | | |
| Infant mortality (0/1) | 1,619,483 | 0.090 | 0.287 | 50,994 | 0.300 | 0.458 | 537,061 | 0.082 | 0.275 | 6,920 | 0.380 | 0.485 |
| Neonatal mortality (0/1) | 1,619,483 | 0.041 | 0.199 | 50,994 | 0.193 | 0.394 | 537,061 | 0.050 | 0.218 | 6,920 | 0.287 | 0.453 |
| Postneonatal mortality (0/1) | 1,552,795 | 0.051 | 0.220 | 41,175 | 0.133 | 0.340 | 510,302 | 0.034 | 0.181 | 4,932 | 0.130 | 0.337 |
| Child mortality (0/1) | 1,473,364 | 0.073 | 0.260 | 35,686 | 0.105 | 0.306 | 492,928 | 0.033 | 0.180 | 4,289 | 0.049 | 0.217 |

Table 3: Mortality rates of boys and girls in different age intervals in sub-Saharan Africa and India

| | Sub-Saharan Africa | | | | India | | | |
|---|---|---|---|---|---|---|---|---|
| | Boys | | Girls | | Boys | | Girls | |
| | Mean | S.D | Mean | S.D | Mean | S.D | Mean | S.D |
| **Infant mortality** | | | | | | | | |
| All twins | 0.323 | 0.468 | 0.277 | 0.447 | 0.393 | 0.489 | 0.366 | 0.482 |
| Male-female twins | 0.307 | 0.461 | 0.280 | 0.449 | 0.338 | 0.473 | 0.348 | 0.476 |
| **Neonatal mortality** | | | | | | | | |
| All twins | 0.211 | 0.408 | 0.174 | 0.379 | 0.308 | 0.462 | 0.265 | 0.441 |
| Male-female twins | 0.202 | 0.401 | 0.180 | 0.384 | 0.260 | 0.439 | 0.251 | 0.434 |
| **Postneonatal mortality** | | | | | | | | |
| All twins | 0.143 | 0.350 | 0.124 | 0.330 | 0.123 | 0.329 | 0.138 | 0.345 |
| Male-female twins | 0.132 | 0.339 | 0.122 | 0.328 | 0.105 | 0.307 | 0.129 | 0.336 |
| **Child mortality** | | | | | | | | |
| All twins | 0.107 | 0.309 | 0.103 | 0.304 | 0.041 | 0.198 | 0.058 | 0.234 |
| Male-female twins | 0.095 | 0.293 | 0.101 | 0.301 | 0.031 | 0.174 | 0.063 | 0.243 |

Table 4: Linear probability model estimates of sex differences in mortality based on twins data from sub-Saharan Africa and India

|  | Sub-Saharan Africa | | | India | | |
|---|---|---|---|---|---|---|
| **Panel A:** Infant mortality | (I) | (II) | (III) | (IV) | (V) | (VI) |
| Male | 0.047*** | 0.045*** | 0.027*** | 0.027** | 0.027** | -0.010 |
|  | [0.004] | [0.004] | [0.005] | [0.012] | [0.011] | [0.016] |
| # Observations | 50,994 | 50,994 | 50,994 | 6,920 | 6,920 | 6,920 |
| **Panel B:** Neonatal mortality | (I) | (II) | (III) | (V) | (VI) | (VII) |
| Male | 0.037*** | 0.036*** | 0.022*** | 0.043*** | 0.045*** | 0.009 |
|  | [0.003] | [0.003] | [0.004] | [0.011] | [0.011] | [0.013] |
| # Observations | 50,994 | 50,994 | 50,994 | 6,920 | 6,920 | 6,920 |
| **Panel C:** Postneonatal mortality | (I) | (II) | (III) | (V) | (VI) | (VII) |
| Male | 0.018*** | 0.018*** | 0.010** | -0.014 | -0.014 | -0.032** |
|  | [0.003] | [0.003] | [0.004] | [0.010] | [0.010] | [0.015] |
| # Observations | 41,175 | 41,175 | 37,958 | 4,932 | 4,932 | 4,324 |
| **Panel D:** Child mortality | (I) | (II) | (III) | (V) | (VI) | (VII) |
| Male | 0.004 | 0.004 | -0.008* | -0.017*** | -0.016** | -0.031*** |
|  | [0.003] | [0.003] | [0.005] | [0.007] | [0.007] | [0.011] |
| # Observations | 35,686 | 35,686 | 30,176 | 4,289 | 4,289 | 3,418 |
|  |  |  |  |  |  |  |
| Twins fixed effect | NO | NO | YES | NO | NO | YES |
| Controls | NO | YES | NO | NO | YES | NO |

Controls include child's year of birth; mother's characteristics: age at survey, education, and marital status; husband's education; household's characteristics: household size, possession of assets such as car, television, radio, and electricity; and a linear control for year of survey, and country fixed effect. Standard errors are in brackets, and are corrected for clustering of observations within mothers.
* significant at 10%; ** significant at 5%; *** significant at 1%

Table 5: Decomposition of sex differences in mortality into the effects of prenatal environment, child biology and parental preferences based on twins data

|  | Sub-Saharan Africa | | | India | | |
|---|---|---|---|---|---|---|
|  | Sex differences in mortality attributable to: | | | Sex differences in mortality attributable to: | | |
|  | Prenatal environment | Child biology | Parental preferences | Prenatal environment | Child biology | Parental preferences |
| Infant mortality | 0.018 | 0.027 | 0 | 0.037 | 0.027 | -0.037 |
| Neonatal mortality | 0.014 | 0.022 | 0 | 0.036 | 0.022 | -0.013 |
| Postneonatal mortality | 0.008 | 0.010 | 0 | 0.018 | 0.010 | -0.042 |
| Child mortality | 0.012 | -0.008 | 0 | 0.015 | -0.008 | -0.023 |

Prenatal environment refers to factors that are external to a child and that occur before conception. In this sense, prenatal environment precedes intrauterine environment, which we are not directly concerned with.

Figure 1: Sex differences in infant and child mortality in Sub-Saharan Africa and India

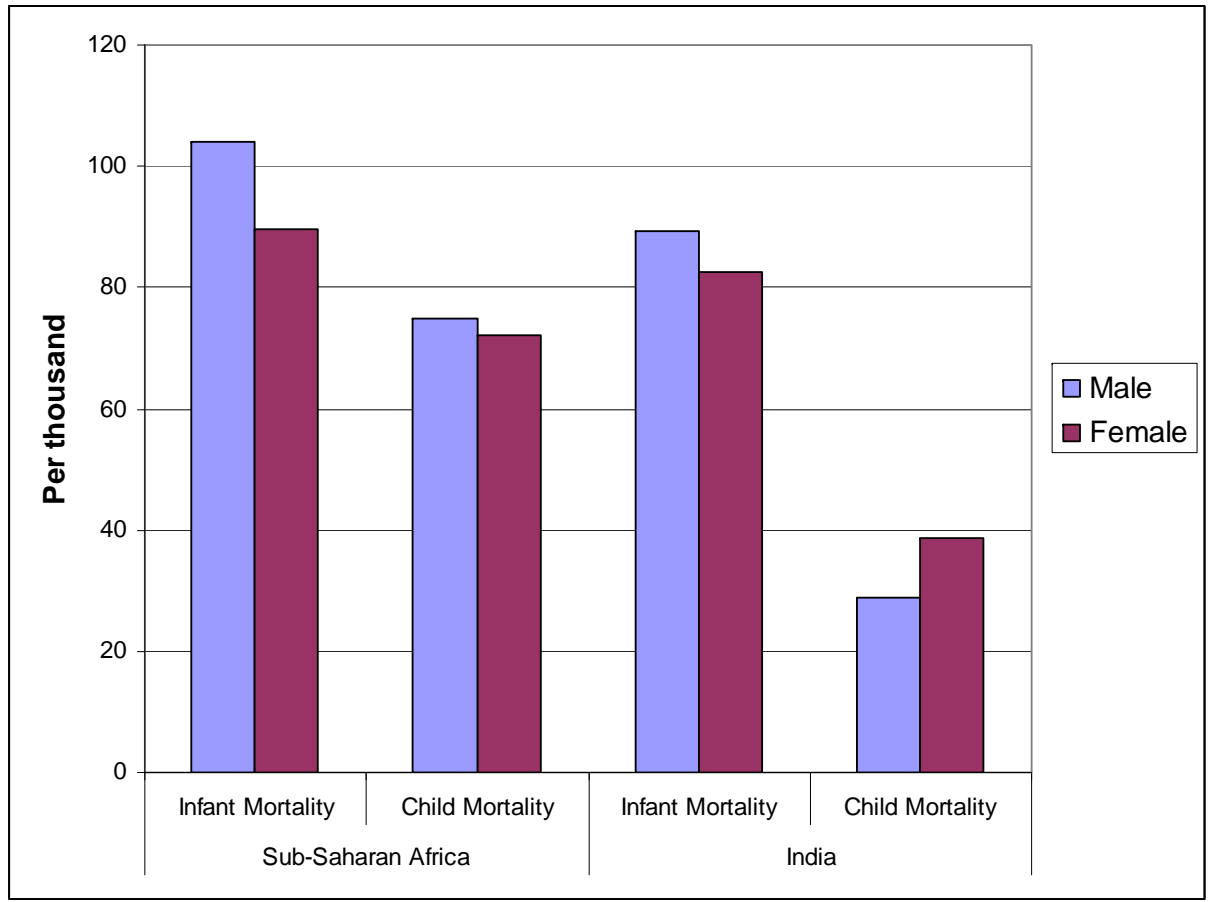

The data are from the Demographic and Health Surveys for sub-Saharan Africa and the National Health Surveys for India. Infant mortality rate is calculated as the probability of dying before the first birthday conditional on being born alive, and child mortality is the probability of dying between the first and fifth birthdays conditional on surviving the first year after birth. We note that mortality rates are greater for males than females during the first year of life in sub-Saharan Africa and India; however, between the first and fifth birthday, while female children continue to die less frequently than their male counterparts in sub-Saharan Africa, the pattern is reversed in India.

**Appendix**

Table A-1: Sample size by country

| Countries | Years of Survey | Total sample size of live births | Sample size of twins | Sample size of singletons |
|---|---|---|---|---|
| India | 1992/93, 1998/99 | 543,981 | 6,920 | 537,061 |
| Sub-Saharan African Countries | | | | |
| Benin | 1996, 2001 | 38,703 | 1,880 | 36,823 |
| Burkina Faso | 1992/93, 1998/99, 2003 | 84,278 | 2,520 | 81,758 |
| Burundi | 1987 | 11,880 | 198 | 11,682 |
| Central African Republic | 1994/95 | 16,933 | 444 | 16,489 |
| Cameroon | 1994, 1998, 2004 | 56,218 | 2,116 | 54,102 |
| Chad | 1996/97, 2004 | 47,175 | 1,350 | 45,825 |
| Comoros | 1996 | 7,907 | 294 | 7,613 |
| Côte d'Ivoire | 1994, 1998/99, 2005 | 45,779 | 1,486 | 44,293 |
| Ethiopia | 2000, 2005 | 84,040 | 1,740 | 82,300 |
| Gabon | 2000 | 16,862 | 532 | 16,330 |
| Ghana | 1988, 1993, 1998, 2003 | 55,743 | 1,890 | 53,853 |
| Guinea | 1999, 2005 | 50,021 | 1,900 | 48,121 |
| Kenya | 1989, 1993, 1998, 2003 | 94,460 | 2,572 | 91,888 |
| Lesotho | 2004 | 14,699 | 422 | 14,277 |
| Liberia | 1986 | 17,261 | 698 | 16,563 |
| Madagascar | 1992, 1997, 2003/04 | 61,362 | 1,282 | 60,080 |
| Malawi | 1992, 1996, 2000, 2004 | 92,571 | 3,584 | 88,987 |
| Mali | 1987, 1995/96, 2001 | 98,535 | 2,788 | 95,747 |
| Mozambique | 1997, 2003 | 63,157 | 2,086 | 61,071 |
| Namibia | 1992, 2000 | 28,309 | 684 | 27,625 |
| Niger | 1992, 1998 | 52,702 | 1,558 | 51,144 |
| Nigeria | 1990, 1999, 2003 | 74,387 | 2,628 | 71,759 |
| Rwanda | 1992, 2000, 2005 | 77,087 | 1,702 | 75,385 |
| Senegal | 1986, 1992/93, 1997, 1999, 2005 | 102,487 | 2,608 | 99,879 |
| South Africa | 1998 | 22,905 | 558 | 22,347 |
| Sudan | 1990 | 25,793 | 684 | 25,109 |
| Tanzania | 1992, 1996, 2004 | 96,491 | 3,228 | 93,263 |
| Togo | 1988, 1998 | 37,009 | 1,532 | 35,477 |
| Uganda | 1988, 1995, 2000/01 | 62,203 | 1,618 | 60,585 |
| Zambia | 1992, 1996, 2001/02 | 70,702 | 2,334 | 68,368 |
| Zimbabwe | 1988, 1994, 1999, 2005/06 | 62,818 | 2,078 | 60,740 |